\documentclass{CHEP2006}
\usepackage{url,xspace,cite}

\author{A.~Buckley\thanks{andy.buckley@durham.ac.uk}, M.~R.~Whalley, W.~J.~Stirling; IPPP, Durham University, England\\
J.~M.~Butterworth, E.~Nurse, B.~Waugh; University College London, England}

\title{HepForge: A lightweight development environment for HEP software}

\DeclareRobustCommand{\etal}{\textit{et al.}\xspace}
\DeclareRobustCommand{\Journal}[4]{{\mbox{\emph{#1}} \textbf{{#2}}, {#3} ({#4})}}
\DeclareRobustCommand{\Rplus}{\protect\nolinebreak\hspace{-.01em}\protect\raisebox{.25ex}{\small\textbf{+}}}
\DeclareRobustCommand{\plusplus}{\Rplus\Rplus}

\DeclareRobustCommand{\cmd}[1]{\texttt{#1}}

\begin{document}

\maketitle


\begin{abstract}
  Setting up the infrastructure to manage a software project can become a task
  as significant writing the software itself. A variety of useful open source
  tools are available, such as Web-based viewers for version control systems,
  ``wikis'' for collaborative discussions and bug-tracking systems, but their use
  in high-energy physics, outside large collaborations, is insubstantial.
  Understandably, physicists would rather do physics than configure project
  management tools.
  
  We introduce the CEDAR HepForge system, which provides a lightweight
  development environment for HEP software. Services available as part of
  HepForge include the above-mentioned tools as well as mailing lists, shell
  accounts, archiving of releases and low-maintenance Web space. HepForge also
  exists to promote best-practice software development methods and to provide a
  central repository for re-usable HEP software and phenomenology codes.
\end{abstract}

\section{Introduction}
In high-energy physics, software development is becoming a discipline in its own
right. The community has benefited from the explosion in popularity of the open
source software (OSS) paradigm and the many utilities developed in that spirit,
with experimental collaborations beginning to deploy systems developed to aid
management of distributed OSS development teams. Such systems include Web-based
bug tracking software, a variety of version control systems and a plethora of
``wiki'' implementations. Various collaborations have deployed these tools to
varying extents.

However, away from the large collaborations, few small research groups have the
resources to investigate, configure and test such tools, which tend to require
expertise in areas such as server configuration which are outside the remit of
most physicists. This need for a substantial initial time-investment, combined
with a reluctance to introduce unfamiliar new working methods, means that uptake
of such systems is sparse among small groups.

CEDAR\cite{cedar:chep04,cedar:web} is just such a small collaboration. In
developing our own software development environment for the JetWeb and HepData
systems, it became evident that such an environment could be extended to set up
well-integrated development tools for an arbitrary number of similar projects
with little replication cost. This system is called
HepForge\cite{cedar:hepforge}, so-named because it provides the same facilities
for particle physicists that the SourceForge.net\cite{sourceforge} service
provides for general open source projects. Additionally, HepForge provides a
convenient forum for central archiving of various re-usable HEP programs.

\section{Introducing HepForge}
\label{sec:hepforge}
HepForge is intended for use by small-to-medium sized HEP projects, specifically
those with the intention to make their code re-usable, portable and documented.
Very specialist code, such as reconstruction software written within
experimental frameworks, is most likely unsuitable and should be maintained
instead within the experiment's code management system. Analysis routines, too,
are typically unsuitable for HepForge: a system specifically designed for them
is the PhyStat\cite{phystat} repository. Project types suited to HepForge would
be, for example, jet clustering algorithms, parton density function (PDF) codes
and Monte Carlo event generators, and indeed projects currently using HepForge's
facilities include the KtJet library\cite{ktjet}, LHAPDF\cite{lhapdf} and a
variety of event generators and related systems. Also suitable, but as yet
unrepresented, would be e.g.  statistics libraries and matrix element
calculation codes.

Aside from the project management facilities provided by HepForge, we hope to
encourage the uptake of more standard configuration and build systems within
high-energy physics. An excellent example of this is the GNU ``autotools'',
comprising \cmd{autoconf}, \cmd{automake} and \cmd{libtool}, which automate many
portability issues in code compilation and library management. These tools are
ubiquitous in the open source world, and the familiarity of their build
procedure is a major boon to developers on OSS projects, but they have yet to be
widely embraced as part of the HEP software process. We hope that through
HepForge more projects can be encouraged and helped in applying such tools to
HEP purposes, if for no other reason than that such standardisation helps to
integrate small tools into larger systems, encouraging code re-use, a
traditional weak area in physics code.

Our intentions explained, we now move on to describe the features offered to
suitable projects by HepForge.

\section{HepForge facilities}
\label{sec:hffacilities}
The main design requirement of HepForge is that the initial learning curve for
users should be very shallow. This includes not forcing a particular working
pattern on the users: they may use as many or as few facilities as they wish,
although the benefits of using multiple features are substantial.

The primary feature of any system for managing software development should be
the facilities provided for managing the code itself. HepForge provides several
such features, of which the first is the provision of the
Subversion\cite{subversion} and CVS\cite{cvs} version control systems.
Subversion is a modern replacement for the well-known CVS system and is highly
recommended as it solves many of CVS's known problems. Both Subversion and CVS
can be accessed via an anonymous read-only method and a read-write developer
mode over SSH, as well as through Web-based viewers. Full-featured shell
accounts can be provided for developers on request and a script is provided to
help in conversion of a CVS repository to the more modern Subversion type.
Additionally, the GNU Arch\cite{arch} and Darcs\cite{darcs} distributed version
control systems are available. Releases of project software can be archived and
are automatically made available, sorted by version, through the HepForge Web
interface.

Project management and source code are linked by the use of the Trac\cite{trac}
bug tracker software, whose primary interface is Web-based. Trac allows projects
to define development milestones, with due dates, and then for bugs (called
``tickets'') to be registered against a particular milestone. The ability of
users to create, modify and close tickets can be controlled by the project
administrator and developers can receive details of how bug fixes are proceeding
by email. A defining feature of Trac is its strong integration with Subversion:
projects using Subversion for their version control can automatically view the
timeline of changes to the code and Subversion commit messages can be used to
make automatic changes to the associated tickets.

Trac additionally provides a ``wiki'' system for collaborative documentation.
Again, users' ability to view, create and edit wiki pages can be set by the
project administrator, and the wiki pages can be referenced in Trac's tickets.
While documentation is usually hailed as important but is nonetheless neglected,
Trac helps to ease this problem by the close integration of the wiki system into
the bug tracker and version control system. While Trac has been integrated
smoothly into the HepForge system, users keen to customise Trac can do so
easily, via a plain text configuration file, templates for Trac's Web pages and
cascading style sheets (CSS)\cite{css}.

As well as providing facilities for managing software development, HepForge
provides Web space for the public presentation of the project. The Web system is
also designed for maximum convenience and is continually evolving new features.
A project named, for example, ``foo'' will automatically have a HepForge Web
page located at \url{http://hepforge.cedar.ac.uk/foo/}, with the full features
of normal HTML and CGI scripts, including PHP and the Spyce inline Python
engine. The displayed name of the project, for example ``Foo'', and meta-data
such as a project description, categories and keywords can be specified via an
XML file: eventually this functionality will be available through the HepForge
Web interface. Project Web pages are processed through a set of custom Apache 2
output filters, which can provide automatic headers and footers on all Web
pages, will correct mistakes in HTML automatically, provide constants support in
CSS files, hide email addresses from search engines and spam harvesters and
allow the page source to be written in more relaxed syntaxes than HTML. Other
filters in development will allow easy automatic highlighting of source code,
inline rendering of LaTeX equations and support for more syntaxes.

Finally, HepForge provides mailing lists for projects: by default a list to
allow users to contact project administrators and a list for announcements to
users are created, but more are available on request. The lists can be
configured by project administrators and list members can manage their own
subscriptions.

From the point of view of the HepForge maintainer, most system tasks are managed
via a carefully designed set of shell scripts, which render the system
maintenance minimal. Common tasks, such as addition of new projects or users, are
highly automated.


For small projects, HepForge offers several benefits over SourceForge and CERN's
deployment of the related Savannah\cite{cernsavannah} system. HepForge provides
full shell accounts and places few restraints on what users may or may not do.
The Web interface, within the project areas, is almost entirely under the
project's control. Useful features such as Subversion support, the highly usable
Trac system and the Web filter system are unique to HepForge. The system is also
in active development and many additional features are planned.

\section{HepCode}
\label{sec:hepcode}
HepForge will eventually be used to provide a final portion of CEDAR, named
HepCode\cite{cedar:hepcode}. This was originally a project to provide access to
well-defined versions of Monte Carlo generator programs, parton distribution
functions and other high-energy physics calculation codes, but with HepForge the
idea has expanded beyond phenomenology codes. In its current state, HepCode is
simply a list of programs, with links to where they can be downloaded:
eventually HepForge will be used to maintain and search this list and to archive
released versions of the code.

\section{Conclusions}
HepForge is a new development environment for re-usable high-energy physics
software. It provides a range of useful facilities with an emphasis on ease of
use and flexibility, and avoids the setup costs that such a system would impose
on small projects. HepForge also aims to encourage the uptake of standard build
systems and development methods in HEP software development.

HepForge currently hosts the core CEDAR projects (JetWeb\cite{cedar:jetweb},
HepML\cite{cedar:hepml}, HZTool\cite{cedar:hztool}, HZSteer\cite{cedar:hzsteer},
Rivet\cite{cedar:rivet}, RivetGun\cite{cedar:rivetgun} and
HepData\cite{cedar:hepdata}) as well as several external projects, including
Herwig\plusplus\cite{herwig++}, ThePEG\cite{thepeg}, Herwig 6\cite{herwig6},
Pythia 6\cite{pythia6}, LHAPDF\cite{lhapdf}, Jimmy\cite{jimmy},
fastNLO\cite{fastnlo}, KtJet\cite{ktjet} and RunMC\cite{runmc}. To date user
feedback has been extremely positive. Other suitable projects are most welcome
to join HepForge.

\section{Acknowledgements}
The CEDAR team would like to thank the UK Particle Physics \& Astronomy
Research Council (PPARC) for their generous support of CEDAR.

\end{document}